\documentclass[aps,prd,superscriptaddress,showpacs,preprint,amsmath,amssymb]{revtex4}
\usepackage{graphicx, bm}
\usepackage{float}
\usepackage[usenames]{color}

\usepackage{subcaption}
\begin{document}

%\tightenlines
\draft

\title{Expected sensitivity on the anomalous quartic neutral gauge couplings in $\gamma\gamma$ collisions at the CLIC}

\author{ A. Guti\'errez-Rodr\'{\i}guez\footnote{alexgu@fisica.uaz.edu.mx}}
\affiliation{\small Facultad de F\'{\i}sica, Universidad Aut\'onoma de Zacatecas\\
         Apartado Postal C-580, 98060 Zacatecas, M\'exico.\\}

\author{E. Gurkanli\footnote{egurkanli@sinop.edu.tr}}
\affiliation{\small Department of Physics, Sinop University, Turkey.\\}

\author{M. K\"{o}ksal\footnote{mkoksal@cumhuriyet.edu.tr}}
\affiliation{\small Department of Physics, Sivas Cumhuriyet University, 58140, Sivas, Turkey.}

\author{V. Ari\footnote{vari@science.ankara.edu.tr}}
\affiliation{\small Department of Physics, Ankara University, Turkey.\\}

\author{ M. A. Hern\'andez-Ru\'{\i}z\footnote{mahernan@uaz.edu.mx}}
\affiliation{\small Unidad Acad\'emica de Ciencias Qu\'{\i}micas, Universidad Aut\'onoma de Zacatecas\\
         Apartado Postal C-585, 98060 Zacatecas, M\'exico.\\}

\date{\today}
%\maketitle

\begin{abstract}
% insert abstract here

The non-Abelian gauge structure of the Standard Model (SM) implies the presence of multi-boson self-interactions. Precise measurements of these interactions allow testing the nature of the SM and new physics contributions arising from the beyond SM. The investigation of these interactions can be approached in a model-independent manner using an effective theory approach, which forms the main motivation of this study. In this paper, we examine the anomalous neutral quartic gauge couplings through the process $\gamma \gamma \rightarrow Z Z$ at the Compact Linear Collider (CLIC) with the center-of-mass energy of $\sqrt{s}=3$ TeV and integrated luminosity of ${\cal L}=5$ $\rm ab^{-1}$.
The anomalous neutral quartic gauge couplings are implemented into FeynRules to generate a UFO module inserted into Madgraph to create background and signal events. These events are passed through Pythia 8 for parton showering and Delphes to include realistic detector effects.
We obtain that the sensitivities on the anomalous quartic neutral gauge couplings with $95\%$ Confidence Level are given as: $f_{T0}/\Lambda^{4}=[-1.06; 1.08]\times 10^{-3}$ ${\rm TeV^{-4}}$, $f_{T1}/\Lambda^{4}=[-1.06; 1.08]\times 10^{-3}$ ${\rm TeV^{-4}}$,$f_{T2}/\Lambda^{4}=[-1.06; 1.08]\times 10^{-3}$ ${\rm TeV^{-4}}$,$f_{T0}/\Lambda^{4}=[-1.06; 1.08]\times 10^{-3}$ ${\rm TeV^{-4}}$, $f_{T5}/\Lambda^{4}=[-4.08; 4.08]\times 10^{-4}$ ${\rm TeV^{-4}}$ and $f_{T8}/\Lambda^{4}=[-1.10; 1.10]\times10^{-4}$ ${\rm TeV^{-4}}$. Our results on the anomalous quartic neutral gauge couplings are set to a more stringent sensitivity concerning the recent experimental limits.

\end{abstract}


\vspace{5mm}

\maketitle

%\narrowtext

\section{Introduction}

To address deficiencies in the Standard Model (SM), such as the asymmetry between matter and antimatter, the strong CP problem, the origin of mass, and the nature of dark energy and dark matter, new physics beyond the Standard Model (BSM) is needed. This exploration examines BSM physics models through various processes at existing and upcoming colliders. One effective approach to search for BSM physics involves investigating anomalous gauge boson couplings, including triple gauge couplings (TGC) and quartic gauge couplings (QGC), which are described by the electroweak $SU(2)_{L}\times U(1)_{Y}$ gauge symmetry of the SM. The measurements of the gauge bosons self-couplings, triple $VVV$ and quartic $VVVV$ with $V= \gamma, Z, W^{\pm}$, provide a fundamental test of the non-Abelian structure of the electroweak theory of the SM and can offer essential information about new physics effects at high energies.
Additionally, knowledge of the electroweak sector of the SM can only be improved through unique direct measurements of the TGC and the QGC. However, the QGC predicted by the SM are $(W^+W^-W^+W^-, W^+W^-ZZ,
W^+W^-\gamma\gamma, W^+W^-Z\gamma)$, while quartic neutral gauge couplings (QNGC) $(ZZZZ,ZZZ\gamma,ZZ\gamma\gamma, Z\gamma\gamma\gamma, \gamma\gamma\gamma\gamma)$ are absent. These couplings can contribute to the production of multiple bosons in colliders, and their precise measurements can either confirm the validity of the SM or lead to the discovery of BSM physics.

By incorporating high-dimensional operators, it is possible to extend gauge boson couplings, thereby affecting the anomalous triple gauge couplings (ATGC) and the anomalous quartic gauge couplings (AQGC). Gauge boson operators have been defined using either linear or non-linear effective Lagrangians. In the non-linear approach, the chiral Lagrangian parameterization is employed to conserve and realize the SM gauge symmetry \cite{1,2}. Within this framework, the ATGC and the AQGC arise from dimension-6 operators. Using dimension-6 operators is particularly convenient for analyzing the AQGC and comparing the results obtained from experiments at the LEP \cite{3}. Additionally, dimension-8 operators are derived using a linear representation of the electroweak gauge symmetry broken by the conventional SM Higgs mechanism. However, with the recent discovery of a new particle consistent with the SM Higgs boson, investigating the AQGC based on the linear effective Lagrangian has gained significant importance.

The CLIC collider offers advantages such as high energy, high luminosity, compact design, and a clean environment \cite{Charles,Aicheler,Blas,Roloff,CLIC-1802-2018,Robson,franc}. Its high energy enables the exploration of new particles and allows for precise measurements. High luminosity increases the production rate of rare events and enhances sensitivity to new physics processes. The compact design makes implementation easier and reduces construction costs. The clean environment, resulting from $e^-e^+$ collisions, enables more precise measurements and better identification of rare processes. Additionally, the CLIC complements the discoveries made at the LHC and the HL-LHC, contributing to a better understanding of the fundamental nature of the universe. The future design of the linear collider is expected to include operation in $e \gamma$ and $\gamma \gamma$ modes. Real photon beams can be generated in these modes by converting the incoming $e^{-}$ and $e^{+}$ beams into photon beams using the Compton backscattering mechanism. The maximum collision energy is anticipated to be $90\%$ for $e \gamma$ collisions and $80\%$ for $\gamma \gamma$ collisions relative to the original $e^-e^+$ collision energy. However, the projected luminosities for $e \gamma$ collisions are $39\%$, while for $\gamma \gamma$ collisions, they are $15\%$ compared to the $e^-e^+$ luminosities \cite{tel}. In this study, our objective is to investigate the sensitivity of the AQGC through the process $\gamma\gamma\rightarrow ZZ$ at the CLIC. The research focuses on a comprehensive physics case for the CLIC with a center-of-mass energy of 3 TeV.

This paper is structured as follows: Section II provides an overview of the dimension-8 operators in the effective field theory. Section III focuses on the process $\gamma\gamma \to ZZ$ and discusses its relevance in the context of the AQGC, referencing recent experimental results. Finally, Section IV presents our conclusions.

\section{Effective field theory approach BSM for the AQGC}

The Effective Field Theory (EFT) approach, described with the SM extended by dimension-8 operators, provides a perfect tool. There are two methods to search for physics BSM. One is to search for new particles. The other is to search for new interactions between known particles. This last method is the one we use to study AQGC. The EFT approach is a simple and elegant way to treat new interactions in BSM. The effective Lagrangian with dimension-8 operators used to investigate the AQGC is given as follows \cite{NPB989-2023,JPG49-2022,EPJC81-2021,JPG47-2020,EPJP135-2020},

\begin{equation}
{\cal L}_{eff}= {\cal L}_{SM} +\sum_{k=0}^1\frac{f_{S, k}}{\Lambda^4}O_{S, k} +\sum_{i=0}^{7}\frac{f_{M, i}}{\Lambda^4}O_{M, i}+\sum_{j=0,1,2,5,6,7,8,9}^{}\frac{f_{T, j}}{\Lambda^4}O_{T, j},
\end{equation}

\noindent where ${\cal L}_{SM}$ denotes the renormalizable SM Lagrange and each $O_{S, k}$, $O_{M, i}$ and $O_{T, j}$ is a gauge-invariant
operator of dimension-8 constructed from SM fields, and $\frac{f_{S, k}}{\Lambda^4}$, $\frac{f_{M, i}}{\Lambda^4}$ and $\frac{f_{T, j}}{\Lambda^4}$ are the corresponding effective Wilson coefficients. According to the structure of the operator, there are three classes of genuine AQGC operators
\cite{Degrande,Eboli3}: independent scalar operators (S-type operators), independent mixed operators (M-type operators), and independent
transverse operators (T-type operators). \\

$\bullet$ Contribution of $O_{S, k}$-type operators:

\begin{eqnarray}
O_{S, 0}&=&[(D_\mu\Phi)^\dagger (D_\nu\Phi)]\times [(D^\mu\Phi)^\dagger (D^\nu\Phi)],  \\
O_{S, 1}&=&[(D_\mu\Phi)^\dagger (D^\mu\Phi)]\times [(D_\nu\Phi)^\dagger (D^\nu\Phi)],  \\
O_{S, 2}&=&[(D_\mu\Phi)^\dagger (D_\nu\Phi)]\times [(D^\nu\Phi)^\dagger (D^\mu\Phi)].
\end{eqnarray}

$\bullet$ Contribution of $O_{M, k}$-type operators:

\begin{eqnarray}
O_{M, 0}&=&Tr[W_{\mu\nu} W^{\mu\nu}]\times [(D_\beta\Phi)^\dagger (D^\beta\Phi)],  \\
O_{M, 1}&=&Tr[W_{\mu\nu} W^{\nu\beta}]\times [(D_\beta\Phi)^\dagger (D^\mu\Phi)],  \\
O_{M, 2}&=&[B_{\mu\nu} B^{\mu\nu}]\times [(D_\beta\Phi)^\dagger (D^\beta\Phi)],  \\
O_{M, 3}&=&[B_{\mu\nu} B^{\nu\beta}]\times [(D_\beta\Phi)^\dagger (D^\mu\Phi)],  \\
O_{M, 4}&=&[(D_\mu\Phi)^\dagger W_{\beta\nu} (D^\mu\Phi)]\times B^{\beta\nu} + {\rm h.c.} ,  \\
O_{M, 5}&=&[(D_\mu\Phi)^\dagger W_{\beta\nu} (D^\nu\Phi)]\times B^{\beta\mu} + {\rm h.c.} ,  \\
O_{M, 7}&=&[(D_\mu\Phi)^\dagger W_{\beta\nu} W^{\beta\mu} (D^\nu\Phi)].
\end{eqnarray}

$\bullet$ Contribution of $O_{T, k}$-type operators:

\begin{eqnarray}
O_{T, 0}&=&Tr[W_{\mu\nu} W^{\mu\nu}]\times Tr[W_{\alpha\beta}W^{\alpha\beta}],  \\
O_{T, 1}&=&Tr[W_{\alpha\nu} W^{\mu\beta}]\times Tr[W_{\mu\beta}W^{\alpha\nu}],  \\
O_{T, 2}&=&Tr[W_{\alpha\mu} W^{\mu\beta}]\times Tr[W_{\beta\nu}W^{\nu\alpha}],  \\
O_{T, 5}&=&Tr[W_{\mu\nu} W^{\mu\nu}]\times B_{\alpha\beta}B^{\alpha\beta},  \\
O_{T, 6}&=&Tr[W_{\alpha\nu} W^{\mu\beta}]\times B_{\mu\beta}B^{\alpha\nu},  \\
O_{T, 7}&=&Tr[W_{\alpha\mu} W^{\mu\beta}]\times B_{\beta\nu}B^{\nu\alpha},  \\
O_{T, 8}&=&B_{\mu\nu} B^{\mu\nu}B_{\alpha\beta}B^{\alpha\beta},  \\
O_{T, 9}&=&B_{\alpha\mu} B^{\mu\beta}B_{\beta\nu}B^{\nu\alpha}.
\end{eqnarray}

\noindent In Eqs. (2)-(19), the different operators are classified regarding Higgs versus gauge boson field content. In this set of genuine
AQGC operators given by Eqs. (2)-(19), $\Phi$ stands for the Higgs doublet, and the covariant derivatives of the Higgs field are given by $D_\mu\Phi=(\partial_\mu + igW^j_\mu \frac{\sigma^j}{2}+ \frac{i}{2}g'B_\mu )\Phi$, and $\sigma^j (j=1,2,3)$ represent the Pauli matrices,
while $W^{\mu\nu}$ and $B^{\mu\nu}$ are the gauge field strength tensors for $SU(2)_L$ and $U(1)_Y$.

In our paper, we examine the sensitivities on the anomalous $ f_ {T,i}/\Lambda^4$ ($i=0,1,2,5,6,7,8,9$) couplings defined by dimension-8 operators related to the AQGC via the process $\gamma \gamma \to Z Z$ at the CLIC. The anomalous quartic gauge boson coupling relevant to the process $\gamma \gamma \to Z Z$ is $ZZ\gamma \gamma$. Operators for these coupling are given by

\begin{eqnarray}
V_{ZZ\gamma \gamma,1}=F^{\mu\nu}F_{\mu\nu}Z^{\alpha}Z_{\alpha},
\end{eqnarray}
\begin{eqnarray}
V_{ZZ\gamma \gamma,2}=F^{\mu\nu}F_{\mu\alpha}Z_{\nu}Z^{\alpha},
\end{eqnarray}
\begin{eqnarray}
V_{ZZ\gamma \gamma,3}=F^{\mu\nu}F_{\mu\nu}Z^{\alpha\beta}Z_{\alpha\beta},
\end{eqnarray}
\begin{eqnarray}
V_{ZZ\gamma \gamma,4}=F^{\mu\nu}F_{\nu\alpha}Z^{\nu\alpha}Z_{\mu\nu},
\end{eqnarray}
\begin{eqnarray}
V_{ZZ\gamma \gamma,5}=F^{\mu\nu}F_{\alpha\beta}Z_{\mu\nu}Z^{\alpha\beta},
\end{eqnarray}
\begin{eqnarray}
V_{ZZ\gamma \gamma,6}=F^{\mu\beta}F_{\alpha\nu}Z^{\mu\beta}Z_{\alpha\nu}.
\end{eqnarray}

Here, $Z^{\mu\nu}=\partial^\mu Z^{\nu}-\partial^\nu Z^{\mu}$. The corresponding coefficients of vertices are represented as follows:

\begin{eqnarray}
\alpha_{ZZ\gamma \gamma,1}=\frac{e^2 \bar{\nu}}{16\Lambda^4}\frac{1}{c_{W}s_{W}}(\frac{s_{W}}{c_{W}} f_{M,0}+2\frac{c_{W}}{s_{W}} f_{M,2}-f_{M,4}),
\end{eqnarray}

\begin{eqnarray}
\alpha_{ZZ\gamma \gamma,2}=\frac{e^2 \bar{\nu}}{16\Lambda^4}\frac{1}{c_{W}s_{W}}(\frac{s_{W}}{2c_{W}} f_{M,7}-\frac{s_{W}}{c_{W}} f_{M,1}-2\frac{c_{W}}{s_{W}} f_{M,3}-2f_{M,5}),
\end{eqnarray}

\begin{eqnarray}
\alpha_{ZZ\gamma \gamma,3}=\frac{c_{W}^2 s_{W}^2}{2\Lambda^4}( f_{T,0}+ f_{T,1}-2f_{T,6}+4f_{T,8})+\frac{c_{W}^2 +s_{W}^2}{\Lambda^4} f_{T,5},
\end{eqnarray}

\begin{eqnarray}
\alpha_{ZZ\gamma \gamma,4}=\frac{c_{W}^2 s_{W}^2}{\Lambda^4}( f_{T,2}+4f_{T,9})+\frac{(c_{W}^2 -s_{W}^2)^2}{2\Lambda^4} f_{T,7},
\end{eqnarray}

\begin{eqnarray}
\alpha_{ZZ\gamma \gamma,5}=\frac{c_{W}^2 s_{W}^2}{\Lambda^4}( f_{T,0}+ f_{T,1}-2f_{T,5}+4f_{T,8})+\frac{(c_{W}^2 -s_{W}^2)^2}{2\Lambda^4} f_{T,6},
\end{eqnarray}

\begin{eqnarray}
\alpha_{ZZ\gamma \gamma,6}=\frac{c_{W}^2 s_{W}^2}{2\Lambda^4}( f_{T,2}- 2f_{T,7}+4f_{T,9}).
\end{eqnarray}

The formalism of the AQGC with dimension-8 operators has been widely discussed in the literature with experimental and phenomenological studies \cite{t1,t2,t4,t5,t6,t66,t8,t9,t10,t11,t12,t13,t14,t15,t16,t17,t19,a1,a2,a3,a33,a4,a5,a6,JPG49-2022,EPJC81-2021,a9,a10,a11,a12,a13,a133,
a14,a15,a16,a17,NPB989-2023,JPG47-2020,a20,a21,a22,a23}.
However, the best experimental limits obtained at 95$\%$ Confidence Level (C.L.) on $ f_ {T,i}/\Lambda^4$ ($i=0,2,5,6,7,8,9$) couplings
are examined through the process $pp \to  Z (\to \nu \bar{\nu}) \gamma jj \to  \nu \bar{\nu} \gamma jj$ at the LHC with $\sqrt{s}=13$ TeV with ${\cal L}_{int}=139$ fb$^{-1}$ and the process $pp \to  Z (\to ll) \gamma jj \to ll \gamma jj$ at the LHC with $\sqrt{s}=13$ TeV with ${\cal L}_{int}=35.9$ fb$^{-1}$ \cite{den,den1}. These are given as follows

\begin{eqnarray}
-9.4<f_ {T,0}/\Lambda^4<8.4 \,(\times10^{-2} \,\text{TeV}^{-4}),
\end{eqnarray}

\begin{eqnarray}
-1.2<f_ {T,1}/\Lambda^4<1.3  \,(\times10^{-1}  \,\text{TeV}^{-4}),
\end{eqnarray}

\begin{eqnarray}
-2.8<f_ {T,2}/\Lambda^4<2.8 \,(\times10^{-1} \,\text{TeV}^{-4}),
\end{eqnarray}
\begin{eqnarray}
-8.8<f_ {T,5}/\Lambda^4<9.9 \,(\times10^{-2} \,\text{TeV}^{-4}),
\end{eqnarray}
\begin{eqnarray}
-4<f_ {T,6}/\Lambda^4<4 \,(\times10^{-1} \,\text{TeV}^{-4}),
\end{eqnarray}
\begin{eqnarray}
-9<f_ {T,7}/\Lambda^4<9 \,(\times10^{-1} \,\text{TeV}^{-4}),
\end{eqnarray}
\begin{eqnarray}
-5.9<f_ {T,8}/\Lambda^4<5.9 \,(\times10^{-2} \,\text{TeV}^{-4}),
\end{eqnarray}
\begin{eqnarray}
-1.3<f_ {T,9}/\Lambda^4<1.3 \,(\times10^{-1} \,\text{TeV}^{-4}).
\end{eqnarray}

Table I presents a list of all the genuine AQGC modified by dimension-8
operators. In particular, the $\gamma\gamma \to ZZ$ channel is especially sensitive to the $f_{T, i}/\Lambda^4$, with $i=0, 1, 2, 5, 6, 7, 8, 9$.
For this reason, we consider only these operators in our study. In addition, the limits obtained for the $f_{T,8}/\Lambda^4$ and $f_{T,9}/\Lambda^4$
coupling parameters are of interest because they can be extracted only by studying the production of electroweak neutral bosons. Therefore, we focus on the $f_{T,i}/\Lambda^4$ parameters to examine AQGC couplings through the process $\gamma\gamma \to ZZ$ at the CLIC.

\begin{table} [ht]
\caption{The AQGC altered with dimension-8 operators are shown with X.}
\begin{center}
\begin{tabular}{|l|c|c|c|c|c|c|c|c|c|}
\hline
%\multicolumn{10}{|c|}{S-type operators}\\
{\rm Gauge-invariant operator} & $WWWW$ & $WWZZ$ & $ZZZZ$ & $WW\gamma Z$ & $WW\gamma \gamma$ & $ZZZ\gamma$ & $ZZ\gamma \gamma$ & $Z \gamma\gamma\gamma$ & $\gamma\gamma\gamma\gamma$ \\
\hline
\cline{1-10}
$O_{S0}$, $O_{S1}$                     & X & X & X &   &   &   &   &   &    \\
$O_{M0}$, $O_{M1}$, $O_{M7}$ & X & X & X & X & X & X & X &   &    \\
$O_{M2}$, $O_{M3}$, $O_{M4}$, $O_{M5}$ &   & X & X & X & X & X & X &   &    \\
$O_{T0}$, $O_{T1}$, $O_{T2}$           & X & X & X & X & X & X & X & X & X  \\
$O_{T5}$, $O_{T6}$, $O_{T7}$           &   & X & X & X & X & X & X & X & X  \\
$O_{T8}$, $O_{T9}$                     &   &   & X &   &   & X & X & X & X  \\
\hline
\end{tabular}
\end{center}
\end{table}

\section{The photon-induced production process $\gamma\gamma \to ZZ$}

A feature of present and future $e^+e^-$ colliders is the possibility to transform it to a photon-collider $\gamma\gamma$ through
the process of Compton backscattering of laser light off the high energy electrons \cite{Eboli-1993,Kingman-1993,Moortgat,Ginzburg,Telnov}. For the evaluation of the total cross-section of the $\gamma\gamma \to ZZ$ signal, we define $m_e$ and $E_e$, as the incident electron beam mass and energy; $E_0$ and $E_\gamma$ as the laser photon and the backscattered photon energies; $\sqrt{s}$ as the
center-of-mass energy of $e^+e^-$ collisions; $\sqrt{\hat s}$ as the center-of-mass energy of the backscattered photon; and $m_Z$ as the mass of the produced
$Z$ gauge boson. Therefore, the $\sigma(s)$ cross-section for the pair production of $Z$ gauge bosons at the photon collider is obtained
by convoluting the subprocess cross-section $\sigma(\hat s)$ with the photon luminosity at an $e^+e^-$ linear collider as follows

\begin{equation}
\sigma(s)=\int^{y_{max}}_{2m_{Z}/\sqrt{s}} dz\frac{dL_{\gamma\gamma}}{dz} \sigma(\hat s), \hspace{1cm} \hat s=z^2 s,
\end{equation}

\noindent where $y_{max}=E_\gamma/E_e$, and $\frac{dL_{\gamma\gamma}}{dz}$ is the photon luminosity, which is defined as:

\begin{equation}
\frac{dL_{\gamma\gamma}}{dz}=2s\int^{y_{max}}_{z^2/y_{max}} \frac{dy}{y} f_{\gamma}(y)f_{\gamma}(z^2/y).
\end{equation}

\noindent In Eq. (40), $f_{\gamma}(y)$  correspond to the energy spectrum of Compton backscattered photons:

\begin{eqnarray}
 f_{\gamma}(y)=\frac{1}{g(\zeta)}\Bigl[1-y+\frac{1}{1-y}-
 \frac{4y}{\zeta(1-y)}+\frac{4y^{2}}{\zeta^{2}(1-y)^{2}}\Bigr] ,
\end{eqnarray}

\noindent with

 \begin{eqnarray}
 g(\zeta)=\Bigl(1-\frac{4}{\zeta}-\frac{8}{\zeta^2}\Bigr)\log{(\zeta+1)}+
 \frac{1}{2}+\frac{8}{\zeta}-\frac{1}{2(\zeta+1)^2} ,
 \end{eqnarray}

\noindent and

 \begin{eqnarray}
 y=\frac{E_{\gamma}}{E_{e}} , \;\;\;\; \zeta=\frac{4E_{0}E_{e}}{M_{e}^2}
 ,\;\;\;\; y_{max}=\frac{\zeta}{1+\zeta},
 \end{eqnarray}

\noindent where $y$ is the energy fraction transferred from the electron to the photon, and the maximum value of the
$y_{max}=\frac{\zeta}{1+\zeta}$ reaches 0.83 when $\zeta=4.8$, and is when the photon conversion efficiency drops
drastically due to the $e^+e^-$ pair production from the laser photons and the photon backscattering.

\subsection{Generation of Signal and background events}

For the computation of the total cross-section and backgrounds, we have implemented the interaction
terms in the MadGraph5\_aMC@NLO \cite{MadGraph} through Feynrules package \cite{AAlloul} as a Universal FeynRules Output (UFO) module \cite{CDegrande}.
However, Monte Carlo event samples for signal and relevant background processes are generated using MadGraph5\_aMC@NLO and passed through Pythia 8.2 \cite{pyt} to include the initial and final parton showering and fragmentation. Using the CLIC configuration cards, the detector response is simulated using Delphes 3.4.2 \cite{del}.

The tree-level Feynman diagram for the process $\gamma\gamma \to ZZ$ is given in Fig. 1.  $\sigma_{\gamma\gamma \to ZZ}$ is the contribution due to BSM physics, which comes from the AQGC $ZZ\gamma\gamma$. In this article, the primary background sources arising from the SM are $\gamma\gamma \to l^+l^-\nu\bar\nu$ and $\gamma\gamma \to WW\to l\nu_l l\nu_l$ which are similar to the signal.
To analyze the process $\gamma\gamma \to ZZ$, we apply a pre-selection criterion starting from the cuts selected: at least
a pair of leptons of the same flavor with opposite charge $(N_l (e, \mu) >= 2)$ must be present in the event, and the transverse momentum
and pseudo-rapidity of the leading and sub-leading charged leptons must be $p_T^{\ell^1} > 10$ GeV, $p_T^{\ell^2} > 10$ GeV and $|\eta^{\ell^1}| < 2.5$, $|\eta^{\ell^2}| < 2.5$, respectively. The criteria include the presence of a dilepton of the same flavor and opposite charge to reconstruct the $Z$ boson. In addition, in Fig. 2, we observe that the event selection is optimized by imposing the distribution of transverse momentum balance ratio $|E^{miss}_T-p^{ll}_T|/p^{ll}_T < 0.7$ for signal $\gamma\gamma \to ZZ$ with some couplings ($f_ {T,0,2,5,8}/\Lambda^4=3$ ${\rm TeV^{-4}}$)
and relevant background processes. This is because the $|E^{miss}_T-p^{ll}_T|/p^{ll}_T$ distribution for the signal deviates significantly from the corresponding backgrounds for values less than 0.7. Another sensitive and important kinematic variable for our study is the transverse component $a^{ll}_T$. This variable
is equal to $a^{ll}_T=|{\bf p}^{ll}_T\times \hat t|$ and correspond to one of two orthogonal components of $p^{ll}_T$ relative
to the dilepton thrust axis $\hat t$ , where $p^{ll}_T={\bf p}^{l_1}_T+ {\bf p}^{l_2}_T$
and $\hat t=({\bf p}^{l_1}_T - {\bf p}^{l_2}_T)/|{\bf p}^{l_1}_T- {\bf p}^{l_2}_T|$ \cite {NIMA602-2009,PRD91-2015}. Therefore, we must take into account
that in the plane transverse to the beam direction, the di-lepton thrust axis is indicated with $\hat t=({\bf p}^{l_1}_T- {\bf p}^{l_2}_T)/|{\bf p}^{l_1}_T
- {\bf p}^{l_2}_T|$. In this case, Fig. 3 shows that by applying $a^{ll}_T > 50$ GeV, the $\gamma\gamma \to ZZ$ signal can be separated from the relevant backgrounds.
As is given in Fig. 4, choosing $E^{miss}_T > 150$ GeV provides a well separation region between the $\gamma\gamma \to ZZ$ signal
and $\gamma\gamma \to WW\to l^-\nu_l l^+\bar\nu_l$ background with $f_ {T,0,2,5,8}/\Lambda^4=3$ ${\rm TeV^{-4}}$.
It is possible to suppress related backgrounds from the $\gamma\gamma \to ZZ$ signal by limiting
the distance $\Delta R(l_1, l_2) < 3$ between leading and sub-leading charged leptons in the $(\eta-\phi)$ plane in Fig. 5. Finally, the cuts used in our analysis are summarized in Table II.

\begin{table}[ht]
%\centering
\caption{List of selected cuts for the analysis of the process $\gamma\gamma \to ZZ$.}
\label{tab2}
\begin{tabular}{p{0.1cm}p{17cm}}
\hline
\hline
& Definitions \\
\hline
 & Preselection: $N_{\ell\,(e,\,\mu)} >= 2$ and a pair of leptons\\
 & Transverse momentum and pseudo-rapidity of the leading and sub-leading charged leptons: $p_T^{\ell^1} > 10$ GeV, $p_T^{\ell^2} > 10$ GeV and $|\eta^{\ell^1}| < 2.5$, $|\eta^{\ell^2}| < 2.5$\\
& $p_T$ balance: $|E_T^{miss}-p_T^{\ell\ell}|/p_T^{\ell\ell} < 0.7$\\
& Transverse component of $p_T^{\ell\ell}$ with respect to the thrust axis: $a_T^{\ell\ell} > 50$ GeV\\
& Missing energy transverse: $E_T^{miss} > 150$ GeV\\
& Minimum distance between leptons: $\Delta R({\ell_{1},\ell_{2}}) < 3.0$\\
\hline
\hline
\end{tabular}
\end{table}

\subsection{Expected sensitivity on $f_{T,j}/\Lambda^{4}$ via $\gamma\gamma$ colliders}

We use the $\chi^2$ distribution to study the expected sensitivity on $f_{T,j}/\Lambda^{4}$ at $95\%$ C. L. through $\gamma \gamma$ colliders. This distribution is defined as follows

\begin{eqnarray}
\label{eq.14}
\chi^{2}=\sum_{i}^{n_{bins}} (\frac{N_{i}^{SIG}-N_{i}^{B}}{N_{i}^{B}\Delta_{i}})^{2}.
\end{eqnarray}

\noindent Here, $N_{i}^{SIG}$ and $N_{i}^{B}$ are the total number of events of the signal and the total backgrounds, respectively. On the other hand, $\Delta_{i}=\sqrt{\delta_{sys}^{2}+\frac{1}{N_{i}^{B}}}$ is the combined systematic ($\delta_{sys}$) and statistical uncertainties for every bin.

The source of systematic uncertainties is mainly based on the cross-section measurements of signal and background processes and higher-order electroweak corrections, the uncertainty in integrated luminosity, photon identification efficiency, and the uncertainties in the energy-momentum scales and resolutions of the final-state particles.
This study considers that the systematic uncertainty for the examined process varies from $\delta_{sys}=5\%$ to 10$\%$. Also, we assume that one of the $Z$ bosons decays to a pair of charged leptons $(l^+l^-; l=e, \mu)$ and the other decays to a neutrino pair. The best limits obtained on the anomalous quartic couplings for the process $\gamma\gamma \to ZZ$ with $\sqrt{s}= 3$ TeV, ${\cal L}=5$ $\rm ab^{-1}$, and $\delta_{sys}=0\%$ can be approximately improved up to about 1.4 times better than the limits obtained with $\delta_{sys}=10\%$.

Fig. 6 shows the expected sensitivities at $95\%$ C.L. on the AQGC through the process $\gamma\gamma \to ZZ$ at the center-of-mass energy of 3 TeV and luminosity of $5$ ${\rm ab^{-1}}$ without systematic uncertainty for the third stage of the CLIC and the latest experimental results in the literature. As illustrated in this figure, our best limits on the AQGC improve much better than the current experimental limits by up to two orders of magnitude. Also, to better understand the results, Table III represents the expected sensitivities for the Wilson coefficients $f_{T,j}/\Lambda^4$, $j=0-2, 5-9$. Here, any couplings are calculated while fixing the other couplings
to zero. Our best sensitivity limits on the AQGC at the CLIC might reach up to the order of magnitude $O(10^{-4}-10^{-2})$. As can be seen from Eqs. (28) and (30), the limits of $f_{T,0}/\Lambda^4$ and $f_{T,1}/\Lambda^4$ should be the same. This can be easily seen in Table III. Also, $f_{T,5}/\Lambda^4$, $f_ {T,8}/\Lambda^4$ and $f_ {T,9}/\Lambda^4$ couplings have the best sensitivities among the AQGC. These results are two orders of magnitude more sensitive than the latest experimental results reported in the literature. Also, the best limits obtained on the anomalous quartic couplings for the process $\gamma\gamma \to ZZ$ with $\sqrt{s}= 3$ TeV, ${\cal L}=5$ $\rm ab^{-1}$, and $\delta_{sys}=0\%$ can be approximately improved up to about 1.4 times better than the limits obtained with $\delta_{sys}=10\%$.

Effective field theory is only valid on the new physics scale, with no unitarity violation. However, high-dimensional operators with non-zero anomalous quartic gauge couplings can lead to a scattering amplitude that violates unitarity at sufficiently high energy values, the so-called unitarity bound. In Ref. \cite{ebe}, a phenomenological study of the two-to-two scattering of electroweak gauge bosons is carried out to determine the partial wave unitarity constraints on the lowest dimensional effective operators that generate the anomalous quartic gauge couplings but leave the triple gauge couplings unaffected. However, the results for $ f_ {T,i}/\Lambda^4$ ($i=0,2,5,6,7,8,9$) couplings are summarised in Table III of Ref. \cite{ebe}.  As can be seen from Table III in our study, our limits for $f_ {T,i}/\Lambda^4$ indicate that they are reliable concerning the unitary violation reported in Ref. \cite{ebe}.

\begin{table}
\caption{Expected lower and upper sensitivity at $95\%$ C.L. on the AQGC through the process
$\gamma\gamma \to ZZ$ at the CLIC with $\sqrt{s}=3$ TeV, ${\cal L}=5\hspace{1mm}\rm ab^{-1}$ under the systematic uncertainties of $\delta_{sys}=0\%$, $5\%$, and $10\%$ are presented.}
\begin{tabular}{|c|c|c|}
\hline
\hline
Couplings (TeV$^{-4}$) & & ${\cal L}=5$ ab$^{-1}$ \\
\hline
\hline
                      &$\delta_{sys}=0\%$       &$[-1.06;1.08]\times10^{-3}$ \\
$f_{T0}/\Lambda^{4}$  &$\delta_{sys}=5\%$       &$[-1.11;1.13]\times10^{-3}$ \\
                      &$\ \, \delta_{sys}=10\%$ &$[-1.27;1.30]\times10^{-3}$ \\
\hline
                      &$\delta_{sys}=0\%$       &$[-1.06;1.08]\times10^{-3}$ \\
$f_{T1}/\Lambda^{4}$  &$\delta_{sys}=5\%$       &$[-1.11;1.13]\times10^{-3}$  \\
                      &$\ \, \delta_{sys}=10\%$ &$[-1.27;1.30]\times10^{-3}$ \\
\hline
                      &$\delta_{sys}=0\%$       &$[-4.76;4.76]\times10^{-3}$ \\
$f_{T2}/\Lambda^{4}$  &$\delta_{sys}=5\%$       &$[-5.04;5.04]\times10^{-3}$ \\
                      &$\ \, \delta_{sys}=10\%$ &$[-5.67;5.67]\times10^{-3}$ \\
\hline
                      &$\delta_{sys}=0\%$       &$[-4.08;4.08]\times10^{-4}$ \\
$f_{T5}/\Lambda^{4}$  &$\delta_{sys}=5\%$       &$[-4.22;4.22]\times10^{-4}$  \\
                      &$\ \, \delta_{sys}=10\%$ &$[-4.39;4.39]\times10^{-4}$ \\
\hline
                     &$\delta_{sys}=0\%$       &$[-1.30;1.18]\times10^{-3}$ \\
$f_{T6}/\Lambda^{4}$ &$\delta_{sys}=5\%$       &$[-1.37;1.24]\times10^{-3}$ \\
                     &$\ \, \delta_{sys}=10\%$ &$[-1.56;1.39]\times10^{-3}$ \\
\hline
                      &$\delta_{sys}=0\%$       &$[-1.33;1.49]\times10^{-2}$  \\
$f_{T7}/\Lambda^{4}$  &$\delta_{sys}=5\%$       &$[-1.38;1.54]\times10^{-2}$ \\
                      &$\ \, \delta_{sys}=10\%$ &$[-1.45;1.61]\times10^{-2}$ \\
\hline
                      &$\delta_{sys}=0\%$       &$[-1.10;1.10]\times10^{-4}$ \\
$f_{T8}/\Lambda^{4}$  &$\delta_{sys}=5\%$       &$[-1.28;1.28]\times10^{-4}$ \\
                      &$\ \, \delta_{sys}=10\%$ &$[-1.48;1.48]\times10^{-4}$ \\
\hline
                      &$\delta_{sys}=0\%$       &$[-1.07;0.89]\times10^{-3}$  \\
$f_{T9}/\Lambda^{4}$  &$\delta_{sys}=5\%$       &$[-1.08;0.39]\times10^{-2}$  \\
                      &$\ \, \delta_{sys}=10\%$ &$[-1.18;0.47]\times10^{-2}$ \\
\hline
\hline
\end{tabular}
\end{table}

\section{Results and Conclusions}

The study of AQGC, as defined by the $SU(2)_{L}\times U(1)_{Y}$ gauge symmetry within the SM, can confirm the model and provide clues to the existence of BSM at a higher energy scale, which can be parameterized with higher-order operators in an EFT. Additionally, future lepton colliders with high energy, luminosity, and detection expectations can open a wide window for new physics research.
In light of these motivations, we examine the sensitivity of AQGC at CLIC options with $\sqrt{s} =3$ TeV with the corresponding integrated luminosity of 5 $\rm ab^{-1}$ through phenomenological analysis of $\gamma\gamma \to ZZ$ production process taking into account charged lepton decay channel of $Z$ boson. In the study, some kinematic variables are used like angular separation $\Delta R(l_1, l_2)$, missing energy transverse $E^{miss}_T$, $p_{T}$ balance $|E^{miss}_T-p^l_T|/p^{ll}_T$ and $a^{ll}_T$ which plays an essential role for distinguishing the signals from the relevant backgrounds. Apart from the cut-based method, a detector simulation has been performed in this study using the Delphes card. Therefore, our results indicate that the expected sensitivity on the Wilson coefficients $f_{T_j}/\Lambda^4$ at $95\%$ C.L. with $\sqrt{s}=3$ TeV, ${\cal L}=5$ $\rm ab^{-1}$ are two orders better than the latest experimental limits. Finally, we understand that $\gamma\gamma$ collisions at the CLIC will have a great potential to probe the AQGC.

\vspace{1.5cm}
%\newpage

\begin{center}
{\bf Acknowledgements}
\end{center}

A. G. R. and M. A. H. R. thank SNII and PROFEXCE (M\'exico). The numerical calculations reported in this paper were fully performed
at TUBITAK ULAKBIM, High Performance and Grid Computing Center (TRUBA resources).

\vspace{1cm}

%\newpage

\newpage

\begin{figure}[H]
\centerline{\scalebox{0.8}{\includegraphics{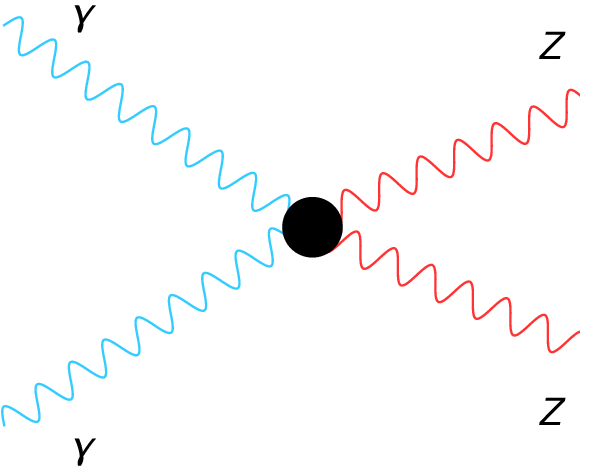}}}
\caption{ \label{fig:gamma} Feynman diagram for the signal process $\gamma\gamma \to ZZ$
induced by the effective $ZZ\gamma\gamma$ vertex. New physics is represented by a black circle in the electroweak
sector can modify the QGC.
}
\end{figure}

\begin{figure}[H]
\centerline{\scalebox{0.6}{\includegraphics{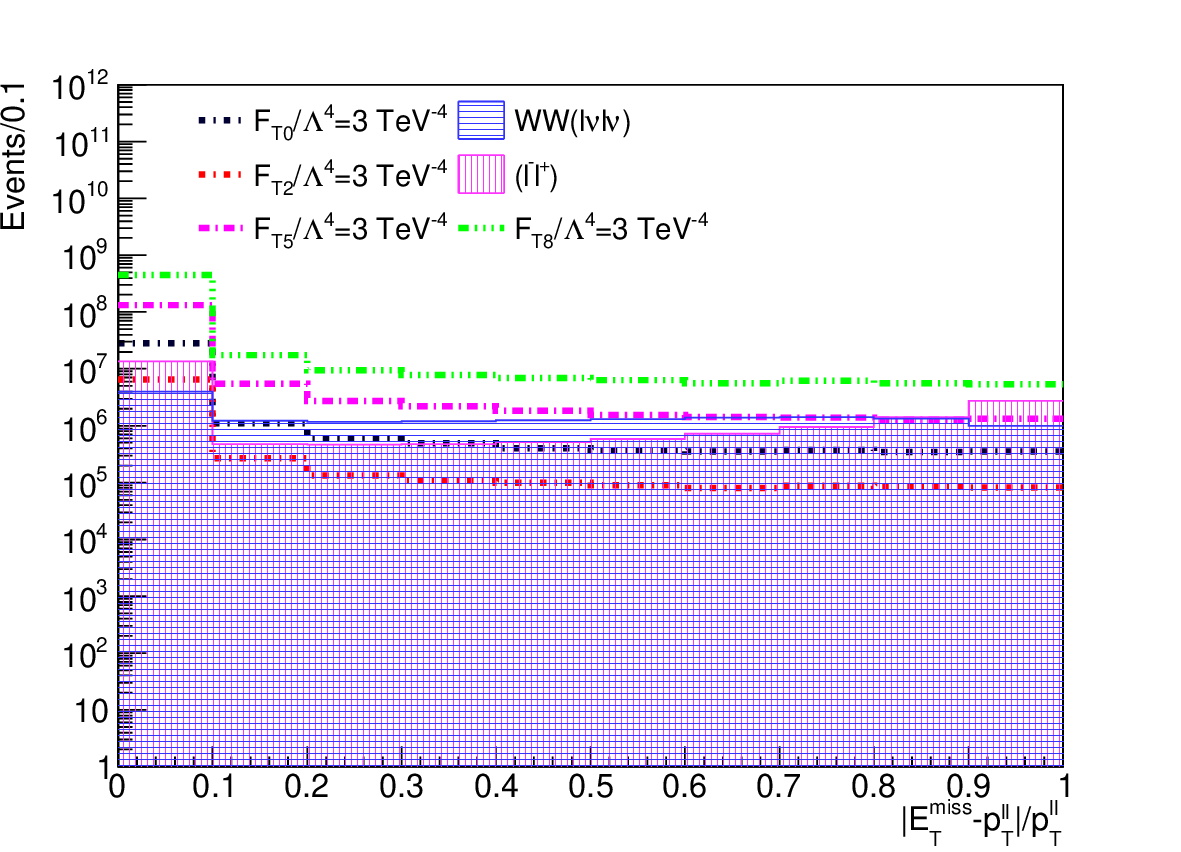}}}
\caption{ \label{fig:gamma} Distribution of transverse momentum balance ratio
$|E^{miss}_T-p^{ll}_T|/p^{ll}_T$ for signal $\gamma\gamma \to ZZ$ with some couplings (
$f_ {T,0,2,5,8}/\Lambda^4=3$ ${\rm TeV^{-4}}$) and relevant background processes.
The dashed lines show the signal predictions for illustrative AQGC parameters.}
\end{figure}

\begin{figure}[H]
\centerline{\scalebox{0.58}{\includegraphics{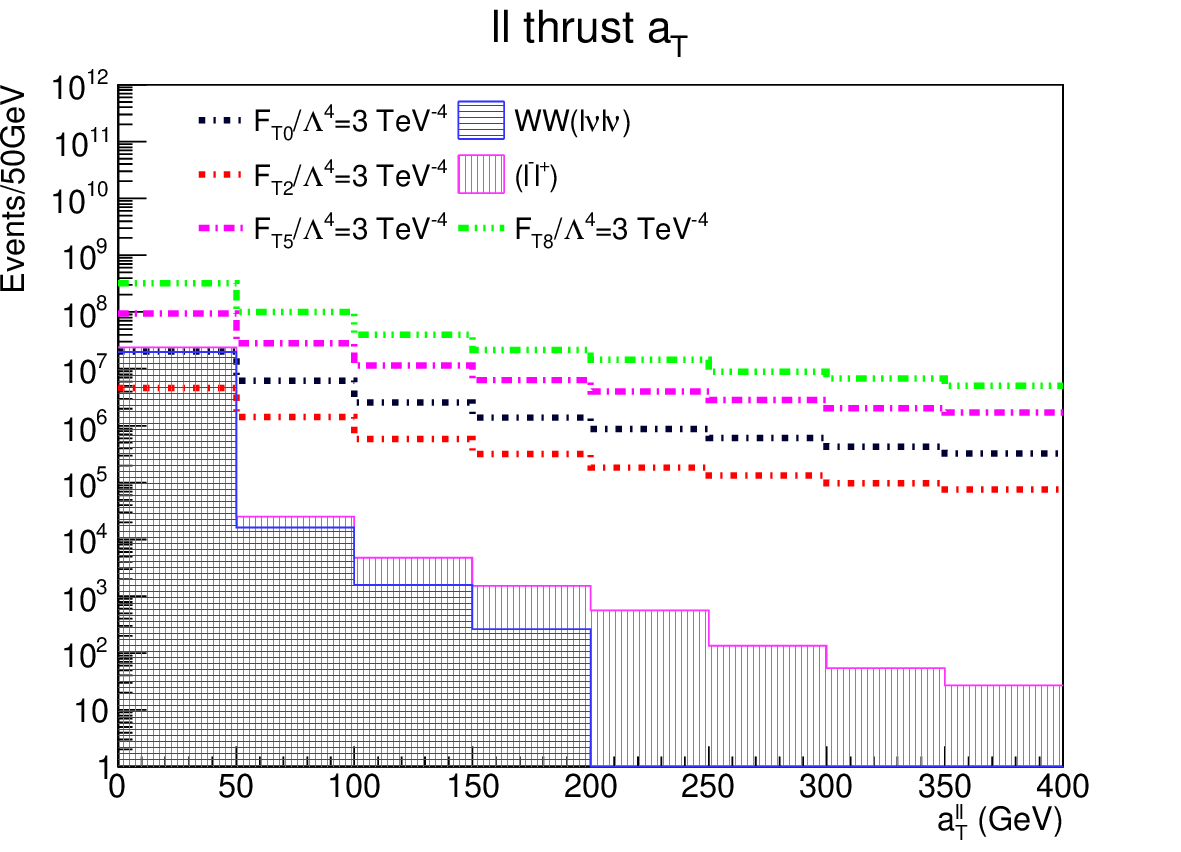}}}
\caption{ \label{fig:gamma} Transverse component $a^{ll}_T$ (GeV) according to the thrust axis for signal
$\gamma\gamma \to ZZ$ with some couplings ($f_ {T,0,2,5,8}/\Lambda^4=3$ ${\rm TeV^{-4}}$) and relevant background processes.
The dashed lines show the signal predictions for illustrative AQGC parameters.}
\end{figure}

\begin{figure}[H]
\centerline{\scalebox{0.58}{\includegraphics{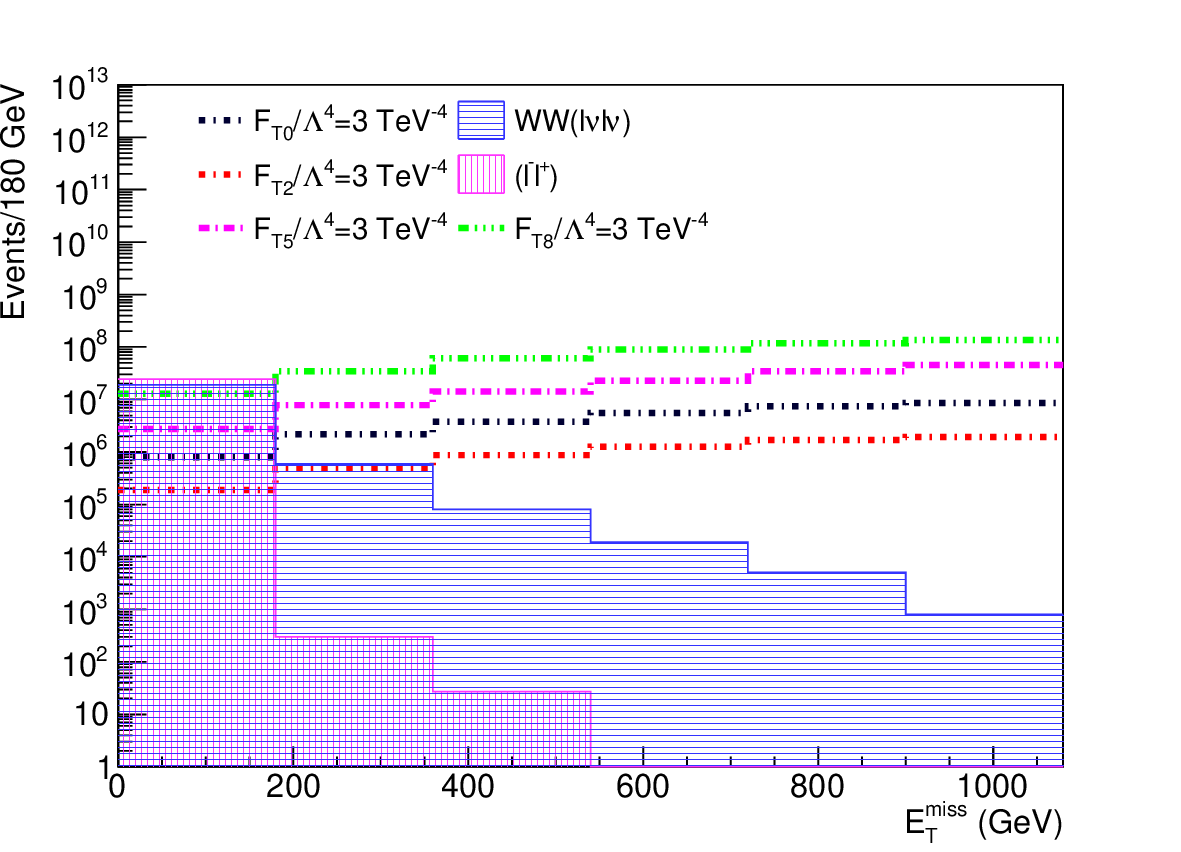}}}
\caption{ \label{fig:gamma} Distribution of missing energy transverse
$E^{miss}_T$ (GeV) for signal $\gamma\gamma \to ZZ$ with some couplings (
$f_ {T,0,2,5,8}/\Lambda^4=3$ ${\rm TeV^{-4}}$) and relevant background processes.
The dashed lines show the signal predictions for illustrative AQGC parameters.}
\end{figure}

\begin{figure}[H]
\centerline{\scalebox{0.58}{\includegraphics{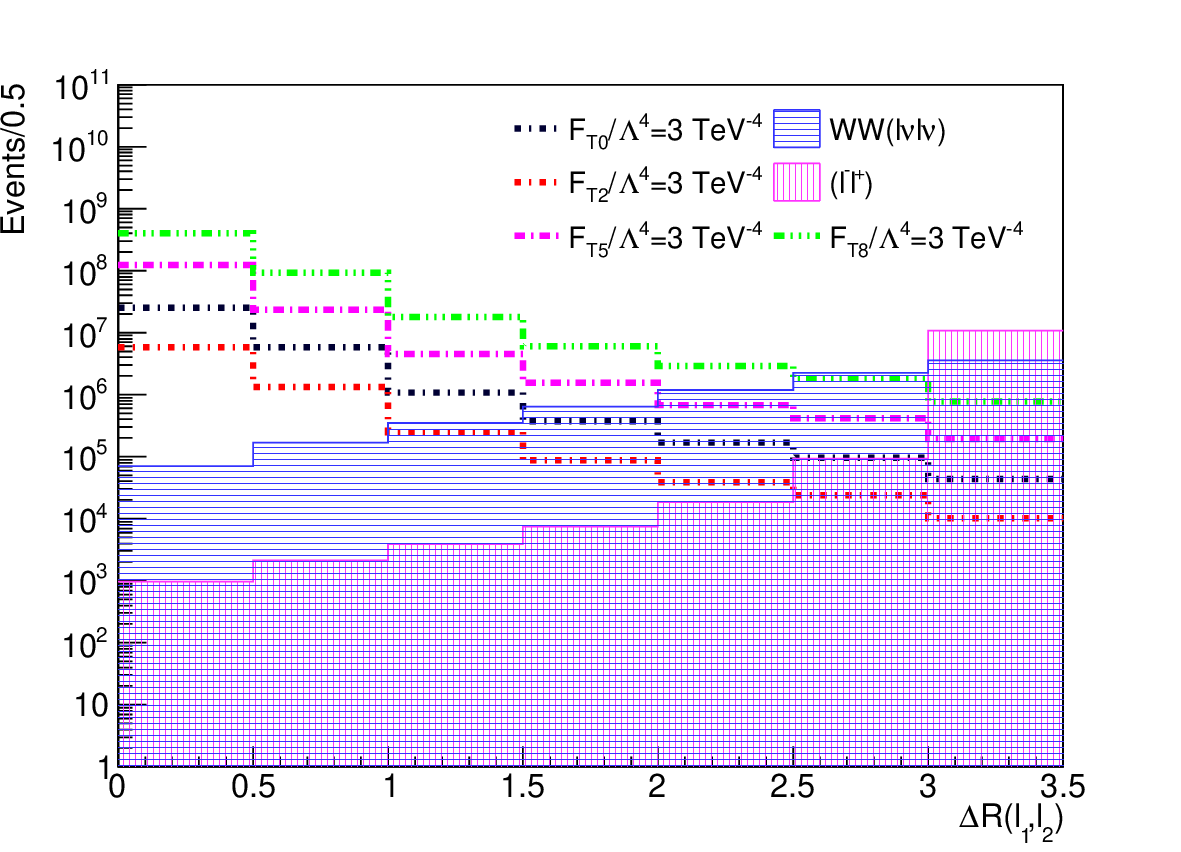}}}
\caption{ \label{fig:gamma} Distance $\Delta R(l_1, l_2)$ between leading and sub-leading charged leptons
for signal $\gamma\gamma \to ZZ$ with some couplings ($f_ {T,0,2,5,8}/\Lambda^4=3$ ${\rm TeV^{-4}}$) and relevant background
processes. The dashed lines show the signal predictions for illustrative AQGC parameters.}
\end{figure}

\begin{figure}[H]
\centerline{\scalebox{0.9}{\includegraphics{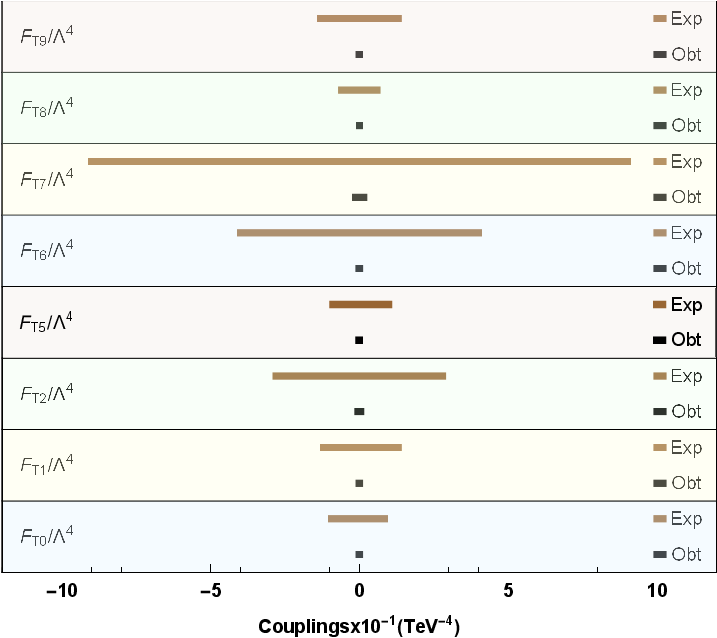}}}
\caption{ \label{fig:gamma} Comparison of the current experimental limits and projected sensitivity
on $f_{T_i}/\Lambda^4$ couplings ($i=0, 1, 2, 5, 6, 7, 8, 9$) at the CLIC with $\sqrt{s}=3$ TeV and luminosity of $5$ ${\rm ab^{-1}}$ without systematic uncertainty.}
\end{figure}

\end{document}